\documentclass[aps,prd,twocolumn,showpacs,amsmath,amssymb, nofootinbib]{revtex4-1}
\newcommand{\BESIIIorcid}[1]{\href{https://orcid.org/#1}{\hspace*{0.1em}\raisebox{-0.45ex}{\includegraphics[width=1em]{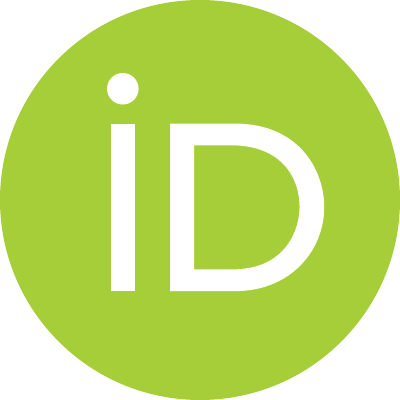}}}}
\usepackage{amsmath}
\usepackage{makecell}
\usepackage{graphicx}
\usepackage{subfigure}
\usepackage{epstopdf}
\usepackage{color}
\usepackage{multirow}
\usepackage{setspace}
\usepackage{overpic}
\usepackage{amssymb}

\usepackage{lineno}
\usepackage{bm}
\usepackage{rotating}
\usepackage[utf8]{inputenc}

\let\oldequation\equation
\let\oldendequation\endequation

\renewenvironment{equation}
  {\linenomathNonumbers\oldequation}
  {\oldendequation\endlinenomath}

\usepackage[bookmarksnumbered, pdfstartview=FitH,colorlinks,urlcolor=blue, citecolor=blue,linkcolor=blue] {hyperref}
\begin{document}

\title{\bf \boldmath
  Measurement of branching fractions of $D^+_s\to K^0_SK^0_S \pi^+\pi^0$ and
  $D^+_s\to K^0_S K^+\pi^0\pi^0$}

\author{
\begin{small}
  \begin{center}
M.~Ablikim$^{1}$\BESIIIorcid{0000-0002-3935-619X},
M.~N.~Achasov$^{4,b}$\BESIIIorcid{0000-0002-9400-8622},
P.~Adlarson$^{79}$\BESIIIorcid{0000-0001-6280-3851},
X.~C.~Ai$^{84}$\BESIIIorcid{0000-0003-3856-2415},
R.~Aliberti$^{37}$\BESIIIorcid{0000-0003-3500-4012},
A.~Amoroso$^{78A,78C}$\BESIIIorcid{0000-0002-3095-8610},
Q.~An$^{75,61,\dagger}$,
Y.~Bai$^{60}$\BESIIIorcid{0000-0001-6593-5665},
O.~Bakina$^{38}$\BESIIIorcid{0009-0005-0719-7461},
Y.~Ban$^{48,g}$\BESIIIorcid{0000-0002-1912-0374}
% ...
\\
\vspace{0.2cm}
(BESIII Collaboration)\\
\vspace{0.2cm} {\it
$^{1}$ Institute of High Energy Physics, Beijing 100049, People's Republic of China\\
% ...
$^{4}$ Budker Institute of Nuclear Physics SB RAS (BINP), Novosibirsk 630090, Russia\\
% ...
$^{37}$ Johannes Gutenberg University of Mainz, Johann-Joachim-Becher-Weg 45, D-55099 Mainz, Germany\\
$^{38}$ Joint Institute for Nuclear Research, 141980 Dubna, Moscow region, Russia\\
% ...
$^{48}$ Peking University, Beijing 100871, People's Republic of China\\
% ...
$^{60}$ Southeast University, Nanjing 211100, People's Republic of China\\
$^{61}$ State Key Laboratory of Particle Detection and Electronics, Beijing 100049, Hefei 230026, People's Republic of China\\
% ...
$^{75}$ University of Science and Technology of China, Hefei 230026, People's Republic of China\\
% ...
$^{78}$ University of Turin and INFN, (A)University of Turin, I-10125, Turin, Italy; (B)University of Eastern Piedmont, I-15121, Alessandria, Italy; (C)INFN, I-10125, Turin, Italy\\
$^{79}$ Uppsala University, Box 516, SE-75120 Uppsala, Sweden\\
% ...
$^{84}$ Zhengzhou University, Zhengzhou 450001, People's Republic of China\\
% ...
$^{\dagger}$ Deceased\\
% ...
$^{b}$ Also at the Novosibirsk State University, Novosibirsk, 630090, Russia\\
% ...
$^{g}$ Also at State Key Laboratory of Nuclear Physics and Technology, Peking University, Beijing 100871, People's Republic of China\\
% ...

}

\end{center}
\end{small}
}

\begin{abstract}
  By analyzing $e^+e^-$ collision data corresponding to an integrated
  luminosity of 7.33~fb$^{-1}$ collected  with the BESIII detector at
  center-of-mass energies ranging from 4.128 to 4.226~GeV, we report the
  observations of the hadronic decays $D^+_s\to K^0_SK^0_S\pi^+\pi^0$ and
  $D^+_s\to K^0_S K^+\pi^0\pi^0$.
  Their decay branching fractions are determined to be
  ${\mathcal B}(D^+_s\to K^0_SK^0_S \pi^+\pi^0)=(4.08\pm0.46_{\rm stat}\pm0.45_{\rm syst})\times 10^{-3}$
  and
  ${\mathcal B}(D^+_s\to K^0_S K^+\pi^0\pi^0)=(3.32\pm0.64_{\rm stat}\pm0.31_{\rm syst})\times 10^{-3}$,
  where the first uncertainties are statistical and the second are systematic.
\end{abstract}

\maketitle

\oddsidemargin  -0.2cm
\evensidemargin -0.2cm

\section{Introduction}
Experimental studies of hadronic $D^+_s$ decays offer important information to
improve the understanding of $CP$ violation and quark SU(3) symmetry breaking
effects in the charm sector~\cite{aboutDsSU31,aboutDsSU32}. Four-body heavy
hadron weak decays provide a rich phenomenology for improving our understanding of
QCD in hadronization, including differential distributions~\cite{Hsiao:2017nga}
and triple-product correlations~\cite{Rui:2021kbn}, and to test \emph{e.g.} the factorization
approach~\cite{Hsiao:2022tfj}. Intensive studies of four-body $D^+_s$ decays
benefit the understanding the $D^+_s$ decay mechanisms. To date, the hadronic
decays $D^+_s\to K^+K^-\pi^+\pi^0$~\cite{DSkkpipi0},
$D^+_s\to K^0_SK^-\pi^+\pi^+$~\cite{DSkskpipi}, and
$D^+_s\to K^0_SK^+\pi^+\pi^-$ have been well investigated in experiments, but
no information about $D^+_s\to K^0_SK^0_S \pi^+\pi^0$ and
$D^+_s\to K^0_S K^+\pi^0\pi^0$ has been reported~\cite{pdg2024}.

In this article, we report the observation of the decays $D^+_s\to K^0_SK^0_S \pi^+\pi^0$
and $D^+_s\to K^0_S K^+\pi^0\pi^0$, and determine their decay branching
fractions, by analyzing $e^+e^-$ collision data corresponding to an integrated
luminosity of 7.33~fb$^{-1}$ collected with the BESIII detector at
center-of-mass energies ($E_{\rm cm}$) between 4.128 and 4.226 GeV. Charge
conjugated decays, $D^-_s\to K^0_SK^0_S \pi^-\pi^0$ and
$D^-_s\to K^0_S K^-\pi^0\pi^0$, are always included throughout this paper.

\section{BESIII detector and Monte Carlo simulation}\label{data_MC}
The BESIII detector~\cite{Ablikim:2009aa} records symmetric $e^+e^-$ collisions
provided by the BEPCII storage ring~\cite{Yu:IPAC2016-TUYA01} in the
center-of-mass energy range from 1.84 to 4.95~GeV, with a peak luminosity of
$1.1 \times 10^{33}\;\text{cm}^{-2}\text{s}^{-1}$ achieved at
$\sqrt{s} = 3.773\;\text{GeV}$. The cylindrical core of the BESIII detector
covers 93\% of the full solid angle and consists of a helium-based multilayer
drift chamber~(MDC), a time-of-flight system~(TOF), and a CsI(Tl)
electromagnetic calorimeter~(EMC), which are all enclosed in a superconducting
solenoidal magnet providing a 1.0~T magnetic field. The solenoid is supported
by an octagonal flux-return yoke with resistive plate counter muon
identification modules interleaved with steel. The charged-particle momentum
resolution at $1~{\rm GeV}/c$ is $0.5\%$, and the specific ionisation energy loss,
${\rm d}E/{\rm d}x$,
resolution is $6\%$ for electrons from Bhabha scattering. The EMC measures
photon energies with a resolution of $2.5\%$~($5\%$) at $1$~GeV in the
barrel~(end-cap) region. The time resolution in the plastic scintillator TOF
barrel region is 68~ps, while that in the end-cap region was 110~ps. The
end-cap TOF system was upgraded in 2015 using multigap resistive plate chamber
technology, providing a time resolution of 60~ps, which benefits 63\% of the
data used in this analysis~\cite{etof}. Luminosities~\cite{lumi} at each energy
are given in Table \ref{tab:mbc}.

Monte Carlo (MC) simulated data samples produced with a
{\sc geant4}-based~\cite{geant4} software package, which includes the geometric
description of the BESIII detector and the detector response, are used to
determine detection efficiencies and to estimate backgrounds. The simulation
models the beam energy spread and initial state radiation (ISR) in the $e^+e^-$
annihilations with the generator {\sc kkmc}~\cite{ref:kkmc}. The inclusive MC
sample includes the production of $D\bar{D}$ pairs (including quantum coherence
for the neutral $D$ channels), the non-$D\bar{D}$ decays of the $\psi(3770)$,
the ISR production of the $J/\psi$ and $\psi(3686)$ states, and the continuum
processes incorporated in {\sc kkmc}~\cite{ref:kkmc}. All particle decays are
modeled with {\sc evtgen}~\cite{ref:evtgen} using branching fractions either
taken from the Particle Data Group~(PDG)~\cite{pdg2024}, when available, or otherwise
estimated with {\sc lundcharm}~\cite{ref:lundcharm}. Final state radiation from
charged final state particles is incorporated using the {\sc photos}
package~\cite{photos2}. The signal MC samples of
$D^+_s\to K^0_SK^0_S \pi^+\pi^0$ and $D^+_s\to K^0_S K^+\pi^0\pi^0$ are
generated with mixed subdecays with branching fractions quoted from the
PDG~\cite{pdg2024}.

\section{Measurement method}
In $e^+e^-$ collisions taken at $E_{\rm cm}$ between 4.128 and 4.226~GeV, the
$D_s^\pm$ mesons are produced mainly via the process
$e^+e^-\to D_s^{*\pm}D_s^\mp\to \gamma(\pi^0)D_s^+ D_s^-$. We perform this
analysis by using the double-tag (DT) method~\cite{DTmethod, Ke:2023qzc},
which was originally developed by the MARKIII Collaboration and has been widely applied in previous BESIII measurements of $D_s^+$ hadronic decays~\cite{DSkkpipi0,DSkskpipi,BESIII:2024oth}. The $D_s^-$ meson,
which is fully reconstructed via one of sixteen hadronic decay modes, is referred to as
a single-tag~(ST) $D_s^-$ meson. The event, in which the transition
$\gamma(\pi^0)$ from $D_s^{*+}$ and the signal decay can be successfully
selected in the presence of ST $D^-_s$ meson, is called a DT event. For a
specific ST mode $i$, the ST and DT yields observed in data are given by
\begin{equation}
N^{i}_{\rm ST} = 2 N_{D^+_sD_s^{*-}} {\mathcal B}^i_{\rm ST} \epsilon^i_{\rm ST},
\end{equation}
and
\begin{equation}
N^{i}_{\rm DT} = 2 N_{D^+_sD_s^{*-}} {\mathcal B}^i_{\rm ST} {\mathcal B}_{\rm sig}
\epsilon^i_{{\rm ST},D_s^+\to {\rm sig}},
\end{equation}
where $N_{D^+_sD_s^{*-}}$ is the number of $D^+_sD_s^{*-}$ pairs produced in
data, ${\mathcal B}^i_{\rm ST}$ is the branching fraction for the ST mode $i$,
${\mathcal B}_{\rm sig}$ is the branching fraction for
$D_s^+\to K_{S}^{0}K_{S}^{0}\pi^{+}\pi^{0}$ or $D_s^+\to K_S^0K^+\pi^0\pi^0$,
$\epsilon^i_{\rm ST}$ is the efficiency of reconstructing the ST mode
$i$~(called the ST efficiency), and $\epsilon^i_{{\rm ST},D_s^+\to \rm sig}$ is
the efficiency of simultaneously finding the ST mode $i$ and the
$D_s^+\to \rm sig$ decay (called the DT efficiency). The branching fractions of
$D_s^{*-}\to\gamma(\pi^0)D_s^-$ are included in the DT efficiency. Based on
these two equations, the branching fraction of the signal decay is determined
as
\begin{equation}
  \mathcal B_{\rm sig}=\frac{N_{\rm DT}}{N^{\rm tot}_{\rm ST} \cdot \epsilon_{\gamma(\pi^0)\rm sig}\cdot\mathcal (B_{K_S^0})^k(B_{\pi^0})^j},
\label{eq1}
\end{equation}
where $N_{\rm DT}$ and $N^{\rm tot}_{\rm ST}$ are the numbers of the DT events
and the ST $D^-_s$ mesons in data summing over tag modes, respectively;
$\epsilon_{\gamma(\pi^0)\rm sig}=\Sigma_i \frac{N_{\rm ST}^i \cdot \epsilon^i_{\gamma(\pi^0)\rm sig}}{N_{\rm ST}^{\rm tot}}$ is the efficiency of detecting the signal decay in the presence of the ST
$D^-_s$ mesons, which is averaged by the ST $D^-_s$ yields $N_{\rm ST}^i$ for
different tag modes; $B_{K_S^0}$~($B_{\pi^0}$) is the branching fraction of
$K_S^0\to \pi^+\pi^-$~($\pi^0\to \gamma\gamma$) from PDG~\cite{pdg2024}; $k$~($j$)
is the number of $K_S^0$~($\pi^0$) mesons in the final state of DT side.

\section{Single-tag candidates}
The ST $D^-_s$ candidates are reconstructed from sixteen hadronic decay modes
of $K^+K^-\pi^-$, $K^+K^-\pi^-\pi^0$, $\pi^-\pi^+\pi^-$, $K_S^0K^-$,
$K_S^0 K^-\pi^0$, $K^-\pi^+\pi^-$, $K_{S}^{0}K_{S}^{0}\pi^{-}$,
$K_{S}^{0}K^{+}\pi^{-}\pi^{-}$, $K_{S}^{0}K^{-}\pi^{+}\pi^{-}$,
$\eta_{\gamma\gamma}\pi^{-}$, $\eta_{\pi^{+}\pi^{-}\pi^{0}}\pi^{-}$,
$\eta_{\gamma\rho^{0}}^{'}\pi^{-}$, $\eta^{'}_{\eta\pi^+\pi^-}\pi^-$,
$\eta_{\gamma\gamma}\rho^{-}$, $\eta_{\pi^+\pi^-\pi^0}\rho^-$, and
$\eta_{\gamma\gamma}\pi^+\pi^-\pi^-$. Throughout this paper, $\rho$ denotes
$\rho(770)$ and the subscripts of $\eta^{(\prime)}$ denote individual decay
modes that are used to reconstruct the $\eta^{(\prime)}$ candidates.

All charged tracks, except for those from $K^0_S$, must be from a region near the interaction point
defined as $|V_{xy}|<1$ cm, $|V_{z}|<10$ cm, and $|\!\cos\theta|<0.93$, where
$|V_{xy}|$ and $|V_{z}|$ are distances of the closest approach in the
transverse plane and along the MDC axis, respectively, and $\theta$ is the
polar angle with respect to the MDC axis. We identify charged particles by
using the combined $dE/dx$ and TOF information, based on which the confidence
levels for pion and kaon hypotheses are calculated. The charged tracks with
confidence level for the pion~(kaon) hypothesis greater than that for the kaon~(pion)
hypothesis are identified as pion~(kaon) candidates.

The $K_S^0$ candidates are formed with the decays $K^0_S\to \pi^+\pi^-$. The
two charged pions must satisfy the requirements $|V_{z}|<20$~cm and
$|\!\cos\theta|<0.93$. The tracks are assumed to be $\pi^+\pi^-$ without any
particle identification~(PID) requirement, and the invariant mass of the
$\pi^+\pi^-$ combination has to be within $(0.486, 0.510)$ MeV$/c^2$. The two
charged tracks are constrained to originate from a common vertex, which is
required to be away from the interaction point by a flight distance of at least
twice the vertex resolution.

Photon candidates are selected by using the isolated showers measured in the EMC.
To suppress
backgrounds from electronic noise or bremsstrahlung, any candidate shower is
required to start within $[0, 700]$ ns from the event start time. The energy of
each shower in the barrel EMC region~\cite{Ablikim:2009aa} is required to be
greater than~25 MeV; while that in the end-cap EMC region must be larger than
50~MeV. To reject the backgrounds associated with charged tracks, the opening
angle between the shower and the extrapolated direction on the EMC of the
nearest charged track is required to be larger than $10^\circ$.

Candidates for $\pi^0$ and $\eta_{\gamma\gamma}$ are formed from the
$\gamma\gamma$ pairs with invariant masses in the ranges of $(0.115,\,0.150)$
and $(0.500,\,0.570)$\,GeV$/c^{2}$, respectively. To improve momentum
resolution, a one-constraint (1C) kinematic fit on the selected $\gamma\gamma$
pair constrains their invariant mass to the known $\pi^{0}$ or $\eta$
mass~\cite{pdg2024}. Candidates for $\rho^{+(0)}$, $\eta_{\pi^0\pi^+\pi^-}$,
$\eta^\prime_{\eta\pi^+\pi^-}$, and $\eta^\prime_{\gamma\rho^0}$ are formed
from the $\pi^+\pi^{0(-)}$, $\pi^0\pi^+\pi^-$, $\eta\pi^+\pi^-$, and
$\gamma\rho^0$ combinations with invariant masses in the ranges of
$(0.570,\,0.970)~\mathrm{GeV}/c^2$, $(0.530,\,0.570)~\mathrm{GeV}/c^2$,
$(0.946,\,0.970)$~GeV/$c^2$ and $(0.940,\,0.976)~\mathrm{GeV}/c^2$,
respectively. In addition, the minimum energy of the $\gamma$ candidate
decaying from $\eta^\prime_{\gamma\rho^0}$ must be larger than 0.1~GeV.

To veto the soft pions from $D^{*+}$ decays, the momentum of any pion which is
not from $K_S^0$, $\eta$, or $\eta^\prime$ is required to be greater than
0.1~GeV/$c$. To reject the peaking background from $D^-_s\to K^0_S\pi^-$ in the
selection of the tag mode $D^-_s\to \pi^-\pi^+\pi^-$, the invariant mass of any
$\pi^+\pi^-$ combination must be outside the range of
$(0.468, 0.528)$~GeV/$c^2$.

\begin{figure}[htbp]
  \centering
    \includegraphics[width=0.45\textwidth] {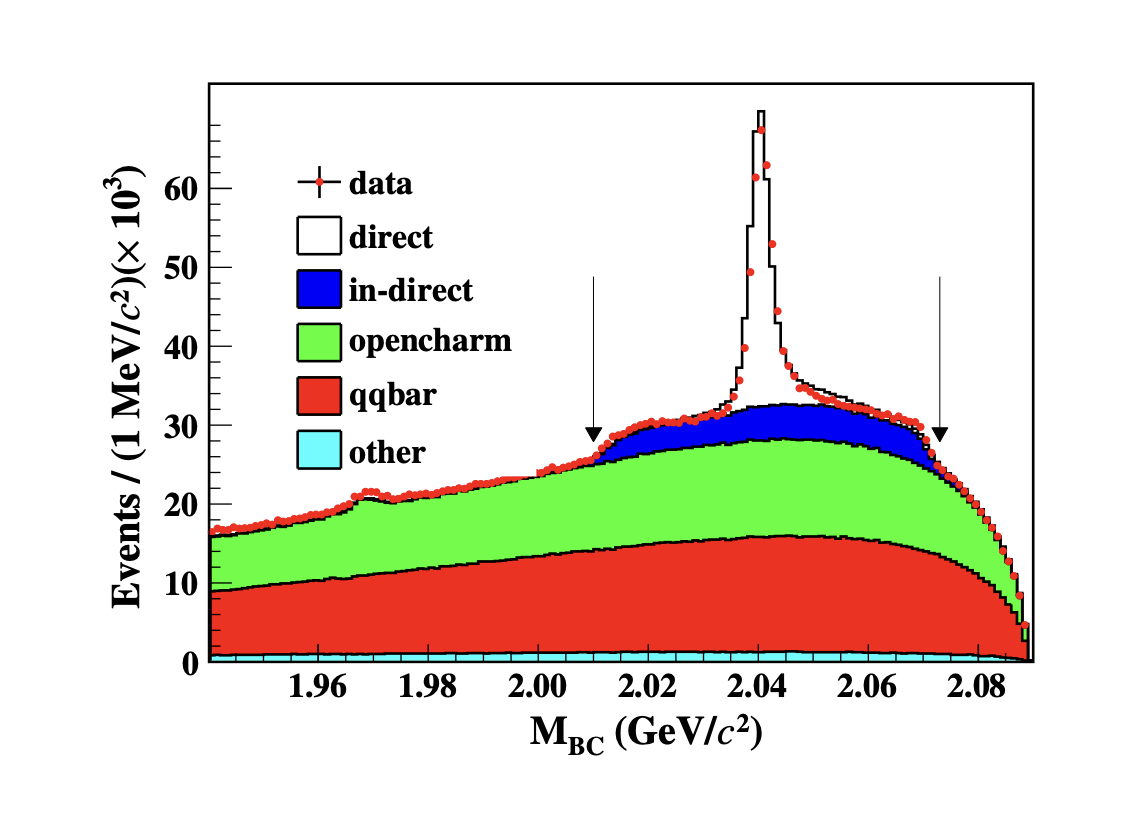}
  \caption{
    Distributions of $M_{\rm BC}$ of the ST $D^-_s$ candidates in data and
    inclusive MC samples at 4.178~GeV. The candidates between the two black
    arrows are retained for further analysis. The colored histograms represent
    various $e^+e^-$ annihilation processes. These histograms are estimated
    with inclusive MC samples mentioned in Sec.~\ref{data_MC}. The
    white~(direct) histogram corresponds to events where the ST $D_s^-$
    candidate is produced directly from the $e^+e^-$ annihilation, while the
    blue~(in-direct) histogram represents events where the ST $D_s^-$ candidate
    originates from the $D_s^*$ decays. The green~(opencharm) and red~(qqbar)
    histograms are the backgrounds caused by $e^+e^-\to D^{(*)}\bar{D}^{(*)}$~($D=D^+, D^0, D_s^+$)
    and $e^+e^-\to q\bar{q}$ processes, respectively. The other~(Cyan)
    histogram includes all other miscellaneous, such as the ISR production of
    $J/\psi$ and $\psi(3686)$, etc.
  }
  \label{fig:diffmbc_ds}
\end{figure}

To suppress the backgrounds from the non-$D_s^{\pm}D^{*\mp}_s$ processes, we
define a kinematic variable, beam-constrained mass of the ST $D_s^-$ candidate,
\begin{equation}
M_{\rm BC}\equiv\sqrt{E^2_{\rm beam}/4c^{4}-|\vec{p}_{\rm tag}|^2/c^{2}}
\end{equation}
where $E_{\rm beam}$ is the beam energy and $\vec{p}_{\rm tag}$ is the momentum
of the ST $D_s^-$ candidate in the rest frame of the $e^+e^-$ system. For each
candidate, the $M_{\rm BC}$ must satisfy the selection criteria listed in
Table~\ref{tab:mbc}. Figure~\ref{fig:diffmbc_ds} shows the $M_{\rm BC}$
distribution of the ST candidates at 4.178~GeV. These requirements retain most
$D_s^-$ and $D_s^+$ mesons from $e^+ e^- \to D_s^{*\mp}D_s^{\pm}$.

\begin{table}[]
  \caption{The integrated luminosity and $M_{\rm BC}$ requirement for each energy~($E_{\rm cm}$) point.}\vspace{0.2cm}
  \begin{tabular}{ccc}
    \hline
    \hline
    $E_{\rm cm}$ (GeV) & Luminosity (pb$^{-1}$) & $M_{\rm BC}$ (GeV/$c^2$) \\\hline
    4.128             &  401.5 & {[}2.010, 2.061{]} \\
    4.157             &  408.7 & {[}2.010, 2.070{]} \\
    4.178             & 3189.0 & {[}2.010, 2.073{]} \\
    4.189             &  569.8 & {[}2.010, 2.076{]} \\
    4.199             &  526.0 & {[}2.010, 2.079{]} \\
    4.209             &  571.7 & {[}2.010, 2.082{]} \\
    4.219             &  568.7 & {[}2.010, 2.085{]} \\
    4.226             & 1091.7 & {[}2.010, 2.088{]} \\
    \hline
    \hline
  \end{tabular}
  \label{tab:mbc}
\end{table}

\begin{figure*}[htbp]
  \centering
  \includegraphics[width=0.9\textwidth]{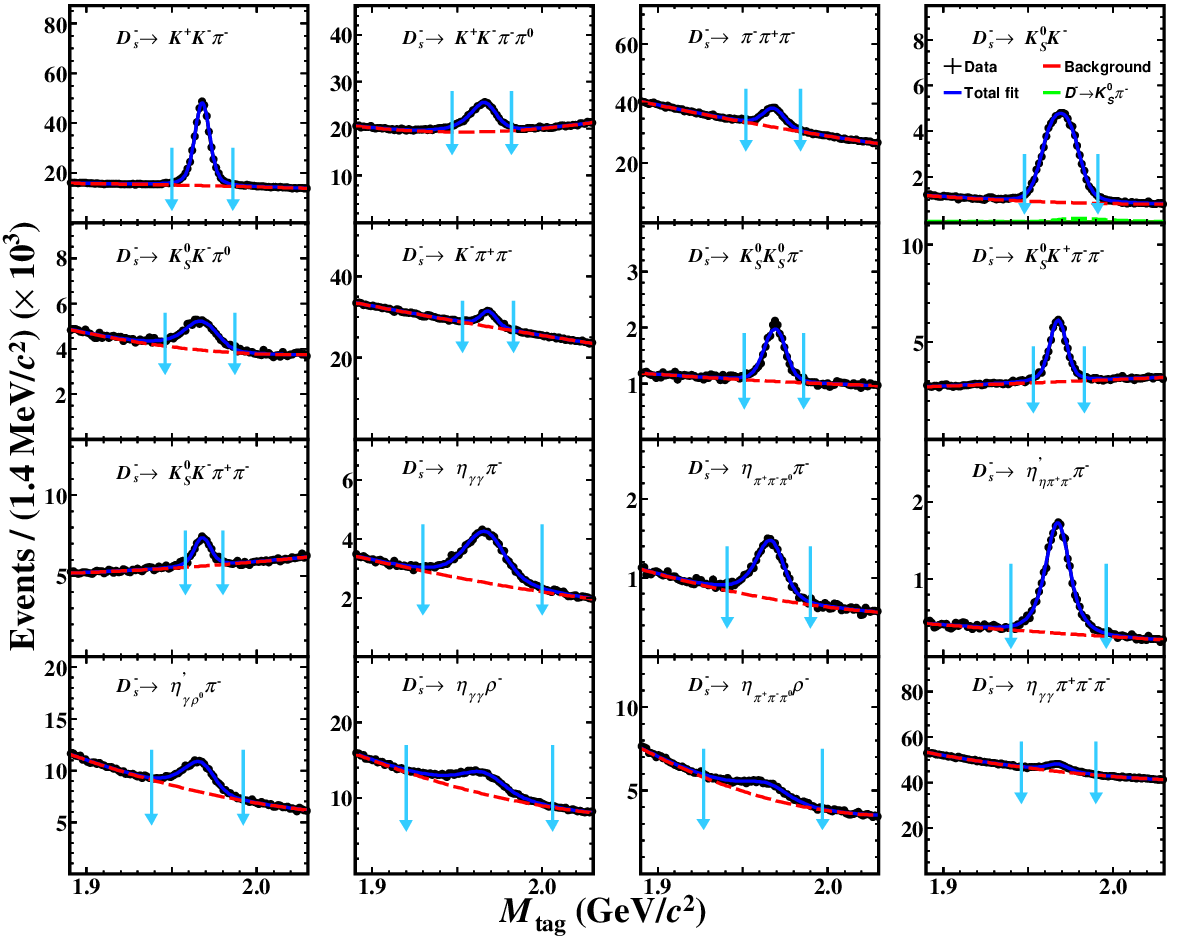} 
  \caption{\footnotesize
    Fits to the $M_{\rm tag}$ distributions of the accepted ST $D^-_s$
    candidates for different tag modes. The points with error bars are data
    summed over all energy points, the blue solid curves are the best fit
    results, and the red dashed curves are the fitted background shapes. For
    the $D^-_s\to K_S^0K^-$ tag mode, the green curve is the
    $D^-\to K_S^0\pi^-$ peaking background. The pair of light blue arrows
    denotes the $M_{\rm tag}$ signal regions.
}
  \label{fig:stfit}
\end{figure*}

If there are multiple candidates present  per tag mode per charge, only the one
with the $D_s^-$ recoil mass
\begin{equation}
  M_{\rm rec} \equiv \sqrt{ \left (E_{\rm cm}\!-\!\sqrt{|\vec p_{\rm tag}|^2\!+\!m^2_{D_s}} \right )^2\!/c^{4}
    -|\vec p_{\rm tag}|^2/c^{2}}
\end{equation}
closest to the known $D_s^{*-}$ mass~\cite{pdg2024} is retained. Here,
$m_{D_s}$ is the nominal $D_s^\pm$ mass~\cite{pdg2024}. The distributions of
the invariant masses ($M_{\rm tag}$) of the accepted candidates for different
tag modes are shown in Fig.~\ref{fig:stfit}. For each tag mode, the yield of
ST $D^-_s$ mesons is obtained from a fit to the individual $M_{\rm tag}$
distribution. In the fit, the signal is modeled by the simulated shape
convolved with a Gaussian function to take into account the resolution
difference between data and simulation. For the tag mode $D_s^-\to K_S^0K^-$,
the shape of the $D_s^-\to K^0_S\pi^-$ peaking background is described by its
simulated shape convolved with the same Gaussian resolution function that is
smeared for the signal shape; while the size of this background is a free parameter in the fit.
The combinatorial background is described by a first to third order Chebychev
function, which has been validated with the inclusive MC sample.
Figure~\ref{fig:stfit} shows the results of the fits to the accepted ST $D^-_s$
candidates from data combined from all energy points. In each plot, the pair of
blue arrows indicates the chosen $M_{\rm tag}$ signal regions. Candidates in the
$M_{\rm tag}$ signal regions are kept for further analysis. By studying the
inclusive MC sample, we find that the process
$e^+e^-\to(\gamma_{\rm ISR})D_s^+D_s^-$ contributes about (0.7-1.1)\% for the
fitted yields of ST $D^-_s$ mesons for different tag modes. For the fitted
yields of the ST $D^-_s$ mesons in data and the ST efficiencies, these
backgrounds have been subtracted away based on inclusive MC simulation. The
$M_{\rm tag}$ requirements, the yields of ST $D^-_s$ mesons ($N_{\rm tag}$) for
different tag modes and the ST efficiencies ($\epsilon_{\rm tag}$) are
summarized in the second and third columns of Table~\ref{tab:bf}, respectively.
In this table, the shown yields of ST $D^-_s$ mesons have been summed over all
energy points, and the shown efficiencies of detecting ST $D^-_s$ mesons are
determined by averaging those by the corresponding yields of ST $D^-_s$ mesons
in data at different energy points.

\begin{figure*}[htbp]
  \centering
  \includegraphics[width=0.8\textwidth]{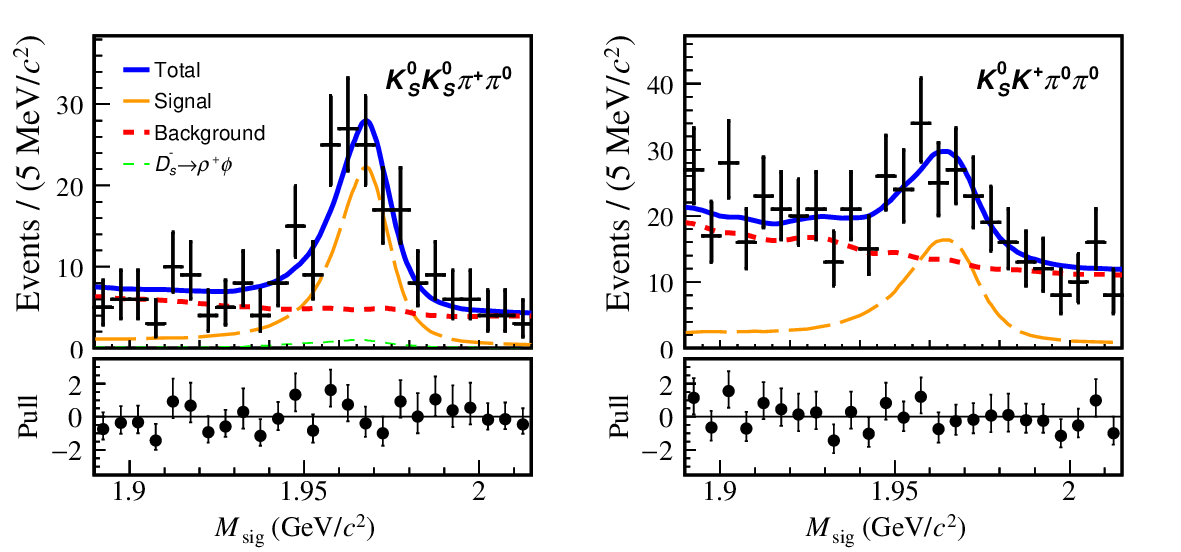} 
  \caption{Fits to the $M_{\rm sig}$ distributions of the accepted DT
    candidates. The points with error bars are data. The blue solid curves are
    the fit results. The yellow and red dashed curves are the fitted
    signals and combinatorial background, respectively.
    The green dashed curve represents
    the background bump caused by $D_s^+ \to \rho^+ \phi$.}
  \label{fig:fit_Umistry1}
\end{figure*}

\begin{figure*}[htbp]
  \centering 
  \includegraphics[width=0.8\textwidth]{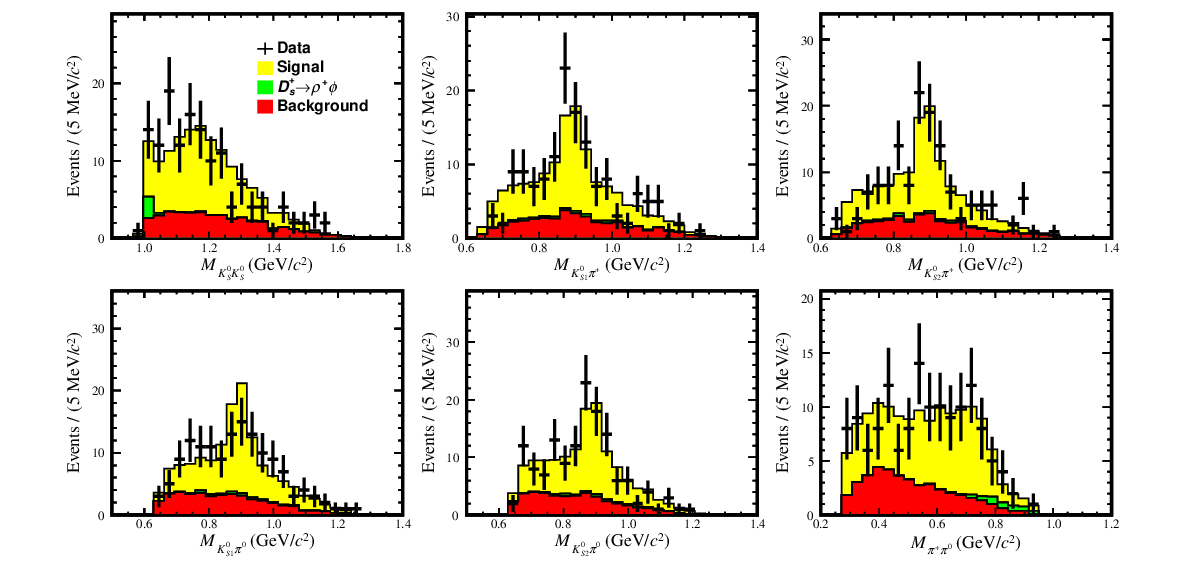}
  \caption{Comparisons of $M_{KK}$, $M_{K\pi}$, and $M_{\pi\pi}$ for
    $D^+_s\to K_{S}^{0}K_{S}^{0}\pi^{+}\pi^{0}$. The points with error bars
    are data, the yellow filled histograms are signal, the green histogram
    represents the peaking background from $D_s^+ \to \rho^+ \phi$,
    and the red filled histograms are simulated backgrounds from the
    inclusive MC sample. Events have been further required to satisfy
    $1.943<M_{\rm sig}<1.983$~GeV/$c^2$.
}
  \label{datamcIM1}
\end{figure*}

\begin{figure*}[htbp]
  \centering
  \includegraphics[width=0.8\textwidth]{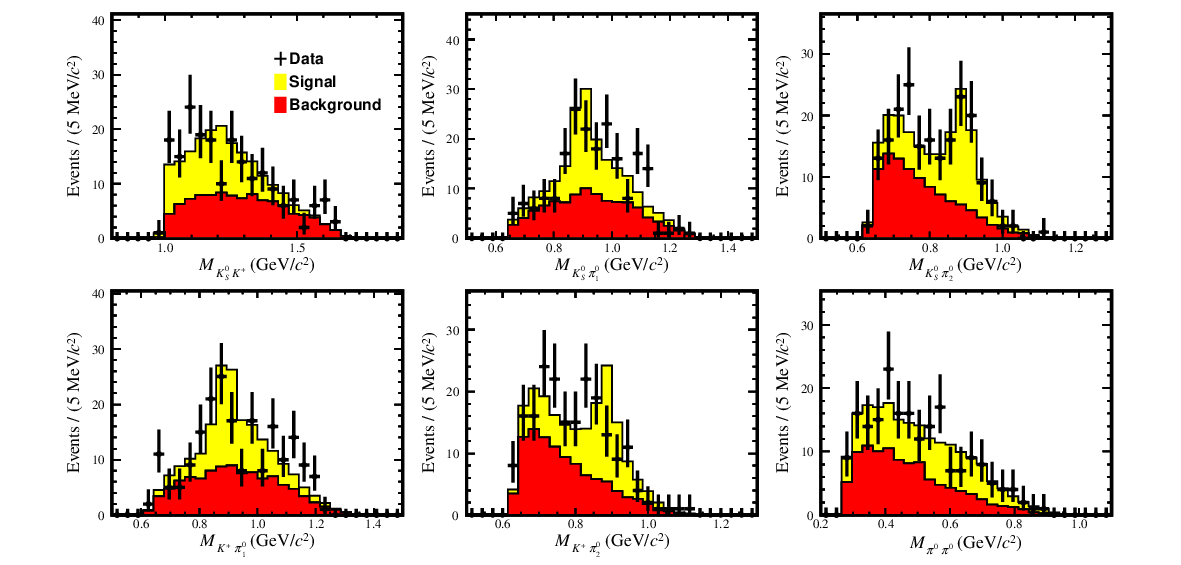}
  \caption{Comparisons of $M_{KK}$, $M_{K\pi}$, and $M_{\pi\pi}$ for
    $D^+_s\to K_S^0K^+\pi^0\pi^0$. The points with error bars are data, the
    yellow filled histograms are signal, and the red filled histograms are
    simulated backgrounds from the inclusive MC sample. Events have been
    further required to satisfy $1.943<M_{\rm sig}<1.983$~GeV/$c^2$.}
  \label{datamcIM2}
\end{figure*}

\begin{table*}[htbp]
  \caption{Requirements of $M_{\rm tag}$, the fitted yields of ST $D^-_s$
    mesons from data~($N_{\rm ST}$), the efficiencies of detecting ST $D^-_s$
    mesons and DT events ($\epsilon_{\rm ST}$, $\epsilon_{\rm DT}$ and
    $\epsilon_{\rm sig}\equiv \epsilon_{\rm DT}/\epsilon_{\rm ST}$) for various
    tag modes, where the uncertainties are
    statistical only, and all efficiencies are in unit of \% and do not
    include the branching fractions of the decays of each daughter particle.}
  \small
  \label{tab:bf}
  \begin{tabular}{l| l| r@{}l r@{}l | r@{}l r@{}l | r@{}l r@{}l}\hline\hline
    Tag mode&$M_{\rm tag}$ (GeV/$c^2$)  &\multicolumn{2}{c}{$N_{\rm ST}$~($\times 10^3$)}&\multicolumn{2}{c}{$\epsilon_{\rm ST}$} & \multicolumn{2}{|c}{$\epsilon_{{\rm DT},K^0_SK^0_S\pi^+\pi^0}$} &\multicolumn{2}{c}{$\epsilon_{{\rm sig},K^0_SK^0_S\pi^+\pi^0}$} & \multicolumn{2}{|c}{$\epsilon_{{\rm DT},K^0_SK^+\pi^0\pi^0}$} &\multicolumn{2}{c}{$\epsilon_{{\rm sig},K^0_SK^+\pi^0\pi^0}$}\\ \hline
    $K^+K^-\pi^-$         &(1.950,1.986)&280.5&$\pm$0.9&40.87  &$\pm$0.04 &3.20&$\pm$0.08 &7.84&$\pm$0.17 &3.10 &$\pm$0.08&7.59&$\pm$0.15 \\
    $K^+K^-\pi^-\pi^0$    &(1.947,1.982)&87.5 &$\pm$0.9 & 11.83&$\pm$0.02 &0.90&$\pm$0.04 &7.60&$\pm$0.34 & 0.71&$\pm$0.04 & 6.01&$\pm$0.25\\
    $\pi^-\pi^+\pi^-$     &(1.952,1.984)&72.7 &$\pm$1.3&51.86  &$\pm$0.09 &3.85&$\pm$0.09 &7.42&$\pm$0.15 &3.70 &$\pm$0.09&7.13&$\pm$0.14\\
    $K_S^0 K^-$           &(1.948,1.991)&63.6 &$\pm$0.5&47.37  &$\pm$0.09 &3.49&$\pm$0.08&7.36&$\pm$0.17&3.30&$\pm$0.08&6.97&$\pm$0.15\\
    $K_S^0 K^-\pi^0$      &(1.946,1.987)&22.2 &$\pm$0.6&17.00  &$\pm$0.07 &1.13&$\pm$0.05&6.67&$\pm$0.24&1.09&$\pm$0.05&6.40&$\pm$0.24 \\
    $K^-\pi^+\pi^-$       &(1.953,1.983)&34.1 &$\pm$0.8&45.41  &$\pm$0.12 &3.32&$\pm$0.08&7.31&$\pm$0.15&3.38&$\pm$0.08&7.44&$\pm$0.15\\
    $K_S^0 K_S^0 \pi^-$   &(1.951,1.986)&10.9 &$\pm$0.2&22.51  &$\pm$0.12 &1.21&$\pm$0.05&5.37&$\pm$0.18&1.26&$\pm$0.05&5.60&$\pm$0.18\\
    $K_S^0 K^+\pi^-\pi^-$ &(1.953,1.983)&29.6 &$\pm$0.4&20.98  &$\pm$0.07 &1.13&$\pm$0.05&5.38&$\pm$0.19&1.16&$\pm$0.05&5.55&$\pm$0.19\\
    $K_S^0 K^-\pi^+\pi^-$ &(1.958,1.980)&15.3 &$\pm$0.4&18.23  &$\pm$0.09 &0.83&$\pm$0.04&4.54&$\pm$0.22&1.09&$\pm$0.05&6.00&$\pm$0.22\\
    $\eta_{\gamma\gamma}\pi^-$&(1.930,2.000)&39.2&$\pm$1.3&48.31&$\pm$0.11&3.42&$\pm$0.08&7.09&$\pm$0.17&3.07&$\pm$0.08&6.36&$\pm$0.15\\
    $\eta_{\pi^+\pi^-\pi^0}\pi^-$&(1.941,1.990)&11.7&$\pm$0.3&23.31&$\pm$0.12&1.61&$\pm$0.06&6.92&$\pm$0.22&1.48&$\pm$0.05&6.36&$\pm$0.22\\
    $\eta'_{\eta\pi^+\pi^-}\pi^-$&(1.940,1.996)&20.0&$\pm$0.2&25.17&$\pm$0.09&1.63&$\pm$0.06&6.48&$\pm$0.20&1.50&$\pm$0.05&5.96&$\pm$0.20\\
    $\eta'_{\gamma\rho}\pi^-$&(1.938,1.992)&49.9&$\pm$1.0&32.45&$\pm$0.08    &2.41&$\pm$0.07&7.43&$\pm$0.19&2.15&$\pm$0.07&6.61&$\pm$0.15\\
    $\eta_{\gamma\gamma}\rho^-$&(1.920,2.006)&77.4&$\pm$1.5&19.92&$\pm$0.04  &1.42&$\pm$0.05&7.12&$\pm$0.25&1.19&$\pm$0.05&5.97&$\pm$0.20\\
    $\eta_{\pi^+\pi^-\pi^0}\rho^-$&(1.927,1.997)&23.4&$\pm$0.5&9.15&$\pm$0.04&0.58&$\pm$0.03&6.35&$\pm$0.33&0.55&$\pm$0.03&6.03&$\pm$0.33\\
    $\eta_{\gamma\gamma}\pi^+\pi^-\pi^-$&(1.946,1.990)&42.7&$\pm$0.3&25.00&$\pm$0.07&1.46&$\pm$0.05&5.84&$\pm$0.20&1.59&$\pm$0.06&6.38&$\pm$0.20\\
    \hline\hline
  \end{tabular}
\end{table*}

\section{double-tag candidates}
In the side recoiling against the tagged $D_s^-$ candidate, the transition
photon or $\pi^0$ and the signal $D^+_s$ decay candidates are selected. We
define the energy difference
$\Delta E \equiv E_{\rm cm} - E_{\rm tag} - E_{\gamma(\pi^0)D^-_s}^{\rm rec} - E_{\gamma(\pi^0)}$,
where
$E_{\gamma(\pi^0)D^-_s}^{\rm rec} \equiv \sqrt{|-\vec{p}_{\gamma(\pi^0)}-\vec{p}_{\rm tag}|^2 + m^2_{D_s}}$, $E_i$
and $\vec{p}_i$ [$i = \gamma(\pi^0)$ or tag] are the energy and momentum of
$\gamma(\pi^0)$ or the tagged $D_s^-$, respectively. In this procedure, we loop over
all unused $\gamma$ or $\pi^0$ candidates and retain the one with the minimum
$|\Delta E|$ for further analysis.

In the selection of the signal candidates, the selection criteria of $\gamma$,
$\pi^0$, $\pi^+$, $K^+$, and $K^0_S$ are the same as those for tag side. For
the $D^+_s\to K^0_SK^0_S \pi^+\pi^0$ ($D^+_s\to K^0_S K^+\pi^0\pi^0$) signal
mode, if there are more than two (one) $K^0_S$ candidates, those with the
longest two (one) decay lengths are retained; if there are more than one~(two)
$\pi^0$ candidate, those with one~(two) minimum $\chi^2_{1\rm C}$ of the 1C
kinematic fit for $\pi^0\to \gamma\gamma$ are kept.

Figure~\ref{fig:fit_Umistry1} shows the $M_{\rm sig}$ distributions of the
accepted candidate events. Unbinned maximum likelihood fits are performed on
these distributions. There is a peaking background caused by $D^+_s\to \rho^+\phi$
for the signal decay $D^+_s\to K^0_SK^0_S\pi^+\pi^0$. In the fits, the signal,
the peaking background and other background shapes are modeled
by the simulated shapes obtained from the signal MC events and the inclusive
MC sample, respectively. The peaking background yield is fixed according to MC
simulation, while the signal and other background yields are set as free
parameters. The significance of
$D^+_s\to K^0_SK^0_S \pi^+\pi^0$ ($D^+_s\to K^0_S K^+\pi^0\pi^0$) 
is estimated to be $8.3\sigma$ ($5.1\sigma$), by comparing the difference in
the likelihoods with and without involving the signal component in the fits,
after taking into account the change of the number of degrees of freedom and
the systematic uncertainties. In addition, we have also examined the
$M_{\rm sig}$ distribution of events in the $K^0_S$ sideband
($0.02 < |M_{\pi^+\pi^-} - M_{K^{0}_{S}}| <  0.044$~GeV/$c^{2}$) of data and
find that they are negligible for both two decays.

The detection efficiencies are estimated to be $(5.94\pm0.01)\%$ and
$(4.49\pm0.01)\%$ for $D^+_s\to K^0_SK^0_S \pi^+\pi^0$ and
$D^+_s\to K^0_S K^+\pi^0\pi^0$, respectively. Figures~\ref{datamcIM1}
and~\ref{datamcIM2} show the comparisons of the invariant masses of two-body
daughter particles of the accepted candidates for
$D^+_s\to K_{S}^{0}K_{S}^{0}\pi^{+}\pi^{0}$ and $D^+_s\to K_S^0K^+\pi^0\pi^0$,
respectively. Good consistency between data and MC simulation ensures the
reliability of the detection efficiencies.

With the known ST yield, the obtained signal yields and the corresponding
signal efficiencies, we obtain the branching fractions of these decays, as
listed in Table~\ref{tab:D0_sigyield}.

\begin{table}[htbp]
  \centering\linespread{1.15}
  \centering
  \caption{The DT yields in data~($N_{\rm DT}$), the signal
    efficiencies~($\epsilon_{\gamma(\pi^0)\rm sig}$), and the obtained
    branching fractions~($\mathcal B_{\rm sig}$), where the uncertainties are
    statistical only.}
  \small
  \begin{tabular}{c|c|c}
    \hline
    \hline
    Signal decay & $D^+_s\to K_{S}^{0}K_{S}^{0}\pi^{+}\pi^{0}$ & $D^+_s\to K_{S}^{0}K^{+}\pi^{0}\pi^{0}$ \\
    \hline
    $N_{\rm DT}$ &$123.7\pm 14.1$&$135.2 \pm 26.2$\\
    $\epsilon_{\gamma(\pi^0)\rm sig}~(\%)$ & $7.22\pm0.01$&$6.79\pm0.01$\\
    $\mathcal B_{\rm sig}~(\times 10^{-3})$ & $4.08\pm0.46$&$3.32\pm0.64$\\
    \hline\hline
  \end{tabular}
  \label{tab:D0_sigyield}
\end{table}

\section{Systematic uncertainty}
The systematic uncertainties in the branching fraction measurements are
discussed below. The uncertainty of the total ST yield, $N_{\rm ST}^{\rm tot}$,
has been assigned as 0.5\% according to Ref.~\cite{NST}. The uncertainties from
tracking and PID of $\pi^+$ or $K^+$ and $\pi^0$ reconstruction are estimated
by using the control samples of $e^+e^-\to K^+K^-\pi^+\pi^-(\pi^0)$ and
$e^+e^-\to 2(\pi^+\pi^-)(\pi^0)$. The systematic uncertainties of
tracking~(PID) efficiencies are assigned as 1.0\%~(1.0\%) per $\pi^+$ or $K^+$.
The photon selection efficiency was previously studied with the
$J/\psi\to\pi^+\pi^-\pi^0$ decays. The systematic uncertainty in the transition
$\gamma(\pi^0)$ reconstruction is assigned as 1.0\%~\cite{gammareconstruction}.
The $\pi^0$ reconstruction efficiencies include photon selection, $\pi^0$ mass
window, and 1C kinematic fit. The difference in efficiencies between data and
MC simulation, 2.0\% per $\pi^0$, is assigned as the systematic
uncertainty~\cite{cite21}.

The systematic uncertainty due to the selection of the best transition
$\gamma/\pi^0$ with the least $|\Delta E|$ has also been studied in
Ref.~\cite{AboutE}. The systematic uncertainty from the selection of the
transition $\gamma\,(\pi^0)$ from $D_s^{*+}$ with the least $|\Delta E|$ method
is estimated by using the control samples of $D^+_s\to K^+K^-\pi^+$ and
$D^+_s\to \eta\pi^0\pi^+$. The difference of the efficiencies of selecting the
transition $\gamma\,(\pi^0)$ candidates between data and MC simulation is taken
as the corresponding systematic uncertainty, which is 0.4\%.

The efficiencies of $K^0_S$ reconstruction, including the tracking efficiencies
of the $\pi^+\pi^-$ pair, decay length requirement, mass window requirement,
vertex fit and second vertex fit, are studied using the control samples of
$J/\psi\to K^0_SK^\mp\pi^\pm$ and $J/\psi\to \phi K^0_SK^\mp\pi^\pm$. The
systematic uncertainty is assigned as the difference of efficiencies between
data and MC simulation, which is 1.5\% per $K^0_S$~\cite{cite22}.

The uncertainties in $M_{\rm sig}$ fit are estimated by comparing the nominal
signal yields with the ones measured by changing the alternative signal and
background shapes. The alternative signal shapes are obtained from the inclusive
MC sample. The alternative background shapes are obtained by varying
the relative fraction of $q\bar q$ background component in the nominal
background shape derived from the inclusive MC sample by $\pm30\%$~\cite{sig15, qq}.
The changes of the fitted signal yields are assigned as the systematic
uncertainties.

The systematic uncertainty of MC generator is estimated by using alternative
signal MC samples. For $D^+_s\to K^0_SK^0_S \pi^+\pi^0$, the known branching
fractions of different sub-resonant decays are varied by $\pm1\sigma$. For
$D^+_s\to K^0_S K^+\pi^0\pi^0$, the average signal efficiency for different
sub-resonant decays is examined. The maximum changes of the signal
efficiencies, 4.9\% and 4.9\%, are assigned as the systematic uncertainties
for $D^+_s\to K^0_SK^0_S \pi^+\pi^0$ and $D^+_s\to K^0_S K^+\pi^0\pi^0$,
respectively.

The uncertainties in the quoted branching fractions of
$\pi^{0} \to \gamma \gamma$ and $K_{S}^{0} \to \pi^+ \pi^-$ are $0.03\%$ and
$0.07\%$. Varying the quoted branching fractions of
$D_s^{*-}\to\gamma(\pi^0) D_s^+$ be $\pm 1\sigma$ affects the DT efficiencies
by 0.2\%, which is assigned as a systematic uncertainty.

The uncertainties due to limited MC statistics are 0.7\% and 0.9\% for
$D^+_s\to K^0_SK^0_S \pi^+\pi^0$ and $D^+_s\to K^0_S K^+\pi^0\pi^0$,
respectively. Due to different reconstruction environments in the inclusive and
signal MC samples, the ST efficiencies determined by the inclusive MC sample
may be different from those by the signal MC sample. This may lead to
incomplete cancellation of the systematic uncertainties associated with the ST
selection, referred to as ``tag bias''. Inclusive and signal MC efficiencies are
compared and the tracking and PID efficiencies for kaons and pions are studied
for different track multiplicities. The resulting ST-average offsets are
assigned as the systematic uncertainties from tag bias and listed in
Table~\ref{tab:totsys}.

\begin{table*}[htbp]
  \centering\linespread{1.15}
	\caption{Systematic uncertainties~(\%) in the branching fraction measurements.}
	\small
{\label{tab:totsys}
   \begin{tabular}{c|c|c}
   \hline\hline
 Source    & $D^+_s\to K_{S}^{0}K_{S}^{0}\pi^{+}\pi^{0}$&$D^+_s\to K_S^0K^+\pi^0\pi^0$\\
   \hline\hline
    $N^{\rm tot}_{\rm ST}$                &0.5  &0.5 \\
    $\pi^{+}$ tracking                    &1.0  &... \\
     $\pi^{+}$ PID                        &1.0  &... \\
     $K^{+}$ tracking                     &...  &1.0 \\
     $K^{+}$ PID                          &...  &1.0 \\
     $\gamma$ and $\pi^0$ reconstruction  &3.0  &5.0 \\
   Best transition $\gamma/\pi^{0}$ selection&0.4  &0.4 \\
    $K_{S}^{0}$  reconstruction           &3.0  &1.5 \\
     Signal shape                         &0.4  &5.4 \\
     Background shape                     &8.8  &1.3 \\
     MC generator                         &4.9  &4.9 \\
     Quoted branching fractions           &0.2  &0.2 \\
     MC statistics                        &0.7  &0.9 \\
      Tag bias & 0.3 & 0.2 \\
\hline
    Total                                 &11.1  &9.2 \\
  \hline
  \hline
  \end{tabular}
}
\end{table*}

For each signal decay, the total systematic uncertainty is obtained by adding
these uncertainties in quadrature. They are assigned to be 11.1\% and 9.2\% for
$D^+_s\to K^0_SK^0_S \pi^+\pi^0$ and $D^+_s\to K^0_S K^+\pi^0\pi^0$,
respectively.

\section{Summary}
By analyzing $e^+e^-$ collision data corresponding to an integrated luminosity
of 7.33~fb$^{-1}$, collected at $E_{\rm cm}$ ranging from 4.128 to 4.226~GeV
with the BESIII detector, the hadronic decays $D^+_s\to K^0_SK^0_S \pi^+\pi^0$
and $D^+_s\to K^0_S K^+\pi^0\pi^0$ are observed for the first time.
The branching fraction of these two decays are determined as
${\mathcal B}(D^+_s\to K^0_SK^0_S \pi^+\pi^0)=(4.08\pm0.46\pm0.45)\times 10^{-3}$
and
${\mathcal B}(D^+_s\to K^0_S K^+\pi^0\pi^0)=(3.32\pm0.64\pm0.31)\times 10^{-3}$,
where the first uncertainties are statistical and the second are systematic.
The measured branching fraction of $D_s^+\to K_{S}^{0}K_{S}^{0}\pi^{+}\pi^{0}$
is consistent with that of $D_s^+\to K_{S}^{0}K^{+}\pi^{0}\pi^{0}$ within
uncertainties.
We notice that the $D_s^+\to \bar K^{*0} K^{*+}$ decay can proceed to decay into the final
states of both $K_{S}^{0}K_{S}^{0}\pi^{+}\pi^{0}$ and $K_{S}^{0}K^{+}\pi^{0}\pi^{0}$.
However, the $D_s^+\to f_0(980)(\to K^0_SK^0_S)\rho^{+}(\to \pi^+\pi^0)$ only decays
into $K_{S}^{0}K_{S}^{0}\pi^{+}\pi^{0}$, but not $K_{S}^{0}K^{+}\pi^{0}\pi^{0}$.
The difference of the two branching fractions may therefore be caused by the contribution from
the $D_s^+\to f_{0}(980)\rho^+$ decay.

Future amplitude analyses of these decays will provide valuable insights into
two-body decay modes involving scalar, vector, axial-vector, and tensor mesons,
thereby improving our understanding of quark-level SU(3)-flavor symmetry.

\textbf{Acknowledgement}

The BESIII Collaboration thanks the staff of BEPCII (https://cstr.cn/31109.02.BEPC) and the IHEP computing center for their strong support. The authors thank Prof. Yu-Kuo Hsiao for helpful discussions. This work is supported in part by National Key R\&D Program of China under Contracts Nos. 2023YFA1606000, 2023YFA1606704; National Natural Science Foundation of China (NSFC) under Contracts Nos. 11635010, 11935015, 11935016, 11935018, 12025502, 12035009, 12035013, 12061131003, 12192260, 12192261, 12192262, 12192263, 12192264, 12192265, 12221005, 12225509, 12235017, 12361141819; the Chinese Academy of Sciences (CAS) Large-Scale Scientific Facility Program; Joint Large-Scale Scientific Facility Funds of the NSFC and CAS under Contract No. U2032104; the Excellent Youth Foundation of Henan Scientific Commitee under Contract No.~242300421044; the Strategic Priority Research Program of Chinese Academy of Sciences under Contract No. XDA0480600; CAS under Contract No. YSBR-101; 100 Talents Program of CAS; The Institute of Nuclear and Particle Physics (INPAC) and Shanghai Key Laboratory for Particle Physics and Cosmology; ERC under Contract No. 758462; German Research Foundation DFG under Contract No. FOR5327; Istituto Nazionale di Fisica Nucleare, Italy; Knut and Alice Wallenberg Foundation under Contracts Nos. 2021.0174, 2021.0299; Ministry of Development of Turkey under Contract No. DPT2006K-120470; National Research Foundation of Korea under Contract No. NRF-2022R1A2C1092335; National Science and Technology fund of Mongolia; Polish National Science Centre under Contract No. 2024/53/B/ST2/00975; STFC (United Kingdom); Swedish Research Council under Contract No. 2019.04595; U. S. Department of Energy under Contract No. DE-FG02-05ER41374

\end{document}